\documentstyle[12pt]{article}

\input{tcilatex}

\begin{document}

\title{On the electron scattering and dephasing by the nuclear spins.}
\author{A. M. Dyugaev \\
L.D. Landau Institute for Theoretical Physics\\
117334 Moscow, Russia. \and I.D. Vagner and P. Wyder \\
Grenoble High Magnetic Field Laboratory \\
Max-Planck-Institut f\"ur Festk\"orperforschung and \\
Centre Nationale de la Recherche Scientifique, \\
BP 166, 38042 Grenoble Cedex 09, France.}
\maketitle

\begin{abstract}
We show that scattering of the conduction electrons by nuclear spins via the
hyperfine interaction may lead the upper limit on the mean free path in
clean metals. Nuclear spins with s \TEXTsymbol{>}1/2 may cause a strong
dephasing in dirty limit due to the quadrupole coupling to the random
potential fluctuations caused by static impurities and lattice imperfections.

{.}
\end{abstract}

\baselineskip4.5ex

Due to the sharply growing interest to the quantum information processing
the study of the electron charge and spin transport in solids has been
refocused to the problems of intrinsic and extrinsic sources of decoherence 
\cite{Esteve1}-\cite{IFS99}. While at low temperatures the phonon scattering
is eliminated, the impurities and electron interactions remain the main
scattering mechanisms \cite{AbrBk}. The magnetic impurities, which can
strongly influence the electron transport, resulting in e.g. Kondo effect,
could be eliminated either by cleaning of the material or by freezing out by
a strong magnetic field. In most of intensively studied conductors there
exist, however, an intrinsic bath of magnetic scatterers: the nuclear spins.
A weak influence of the nuclear spins on the resistivity in strongly doped
bulk semiconductors was reported in \cite{SKL71,GJ96}. A striking example of
their influence on the electron transport in low dimensional semiconductors
is the observation of sharp spikes in magnetoresistance \cite{KDKDWB99}
under the quantum Hall effect conditions.

Much less attention was paid to the magnetic scattering of conduction
electrons by nuclear spins in metals. There is however strong evidence that
conventional scattering can not explain anomalies in residual resistivity at
low temperature \cite{BPS90}. While the normal metals have a quite similar
electronic structure, the experimentally observed temperature dependence of
the dephasing time $\tau _{\varphi }$ may be quite different. This was shown
in very recently in \cite{Esteve1}, where the value of \ $\tau _{\varphi }$
was defined by the magnetoresistance measurements of long metallic wires $%
Cu,Au,Ag$ in a wide temperature interval $10^{-2}<T<10^{^{o}}$K. In $Cu$ and 
$Au$ wires $\tau _{\varphi }$ saturates at low temperatures which 
contradicts the standard theory \cite{AAK82}. Strangely enough the Ag wires
do not show the saturation to the lowest temperatures, in accordance with 
\cite{AAK82}. The possibility of the hyperfine origin of this discrepancy
will be discussed later in this paper.

Here we study the contribution of the hyperfine contact (Fermi) interaction
between the conduction electrons and nuclear spins to the temperature and
magnetic field dependence of resistivity $\rho (T,H)$ . We show that as a
result of electron-nuclear interaction the residual resistivity in
isotopically clean metals is not vanishing even when the impurity
concentration $C_{o}\rightarrow 0$ (the universal residual resistivity,
URR). The space periodicity of nuclei is of no importance, as long as the
nuclear spins are disordered and acts as magnetic impurities with the
concentration $C_{n}\approx 1$ . It follows that in this temperature
interval URR reflects the existence of an upper limit for the mean free path
of conduction electrons. This scattering is not operative at extremely low
temperatures ($T\leq $ $10^{-7}$ $^{o}K$in Cu, for example ) when the
nuclear spins are ordered.

The residual ''nuclear ''resistivity is due to the Fermi (contact) hyperfine
interaction between the nuclear and the conduction electron spins:

\begin{equation}
V_{en}=-\frac{8\pi }3\mu _e\mu _h\Psi _e^2(0)\equiv \mu _nH_e  \label{Ven1}
\end{equation}
here $\mu _e$ and $\mu _h$ are the operators of the electron and nuclear
magnetic moments, $\Psi _e^2(0)\propto $ $Z$ is the value of the conduction
electron wave function on the nuclei with the nuclear charge $Z$ and $H_e$
is the magnetic field induced on nuclei by the electrons.

Let us estimate $V_{en}$ . In atomic units: $\hbar =m_e=e=1$

\begin{equation}
V_{en}\approx Z\alpha ^2\frac{m_e}{m_n}Ry  \label{Ven2}
\end{equation}
where $m_e,m_h$ are the electron and the nucleon masses, respectively; $%
Ry=27 $ ev and $\alpha =\frac 1{137}$ is the fine structure constant.

In metals the effective electron-nuclear interaction constant is

\begin{equation}
g_{n}\equiv \frac{V_{ne}}{\epsilon _{F}}\approx 10^{-7}Z\frac{Ry}{\epsilon
_{F}}  \label{gn}
\end{equation}
where the Fermi energy $\epsilon _{F}$ varies in wide interval $\left(
0.01\div 1\right) Ry$.The interaction constant $g_{n}$ varies from is $%
10^{-6}$ for $Li$ to $10^{-1}$ in doped semiconductors with low $\epsilon
_{F}$ . This estimate of $g_{n}$ is in a good agreement with the
experimentally observed values of $H_{e}$ on the nuclei \cite{RadoSuhlIIA}.

The total residual resistivity is therefore a sum of the impurity $\rho
_o(T\rightarrow 0)\sim C_o$ and the nuclear spin $\rho _n(T\rightarrow
0)\sim g_n^2$ contributions: 
\[
\rho _o^{+}(0^{+})\approx \rho _{oo}(C_o+g_n^2) 
\]
which follows also from calculations, based on the magnetic impurity
scattering technics introduced in \cite{Abr65} . Here $0^{+}$ is the limit $%
T\rightarrow 0$ , while $T\gg T_c$,where $T_c$ is the temperature of the
nuclear ordering, and $\rho _{oo}\approx 1$ in atomic units:$\rho
_{oo}\approx 10^{-17}$ sec \cite{AbrBk,AMBk}. The nuclear contribution to
resistivity starts to be operative when the impurity concentration is $C_o
\sim g_n^2$ .

In the limit of an ideally pure ($C_o=0$) metal the universal residual
resistivity $\rho_{URR}$is, therefore $\rho_{URR}\geq \rho_{oo}g_n^2$ and
the mean free path is limited by $\frac{10^{-8}}{g_n^2}$ cm. This yields $%
10^4$ cm in $Li$ and $10^{-2}$ cm for the rear earth metals. It is
interesting to note that in materials with even-even nuclei (zero spin) ,
like in $Ca$ ,$Ni$ , $Fe$, $Ce$ and isotopically clean graphite $C$ , where
the electron-nuclear scattering is absent, the URR would not be observed.

Consider now the contribution to the temperature and magnetic field
dependence of the residual resistivity caused by the hyperfine interaction
between the conduction electrons and the nuclear spins. The temperature and
the magnetic field dependence of residual resistivity due to nonmagnetic
impurities is due mostly to the mesoscopic effects, and is vanishing in the
limit $C_{o}\rightarrow 0$ \cite{ALW91} . In a magnetic field such that $\mu
_{e}H\gg T$ the magnetic impurities freeze out and the Kondo effect is
quenched. In order to freeze out the nuclear spins however one should apply
much higher magnetic fields, $\mu _{n}H\gg T$ . Therefore in the temperature
interval $\mu _{e}H\gg T\gg $ $\mu _{n}H$ \ the nuclear spin contribution
may prevail in metals with magnetic impurities.

The temperature and the magnetic field dependence of the electron-nuclear
scattering contribution to resistivity can be written as

\begin{equation}
\rho _{n}(T)=\rho _{n}(\infty )f_{n}(x)  \label{ron1}
\end{equation}
where $x=\frac{\mu _{n}H}{T}$ and the asymptotic of \ the function $f_{n}(x)$
is given in \cite{DVW96}. Nuclei with spin $\frac{1}{2}$ in a magnetic field
are equivalent to a two-level system and the function $f_{n}(x)$ can be
defined analytically by methods developed in \cite{AGD} to be: 

\begin{equation}
f_{n}(x)=\frac{2x}{sh2x}  \label{fn3}
\end{equation}

In the limit $T\gg \mu _{n}H$ the temperature dependent part of $\rho $ is

\begin{equation}
\rho (T)-\rho (0^{+})\cong \rho _{00}\left( \frac{T^{2}}{\varepsilon _{F}^{2}%
}-g_{n}^{2}\frac{(\mu _{n}H)^{2}}{T^{2}}\right)   \label{ro5}
\end{equation}
Since the recent experimental data are plotted as $\frac{\partial \rho }{%
\partial T}$ versus $T$ \cite{BPS90} we note that the derivative $\frac{%
\partial \rho }{\partial T}$ experiences a minimum at $T\sim \sqrt{%
g\varepsilon _{F}\mu _{n}H}.$

In metals like $Li,Na,K,Rb,Cs,Au,Cu,Al,In$ the nuclear magnetic moments $%
I\neq \frac{1}{2}$ and even without external magnetic field their $2I+1$
degeneracy is lifted partially by the quadrupole effects (in the case of
cubic crystal symmetry the quadrupole splitting of the nuclear levels may
happen due to the defects \cite{WintBk},\cite{TB68},\cite{AR59}) . The
hyperfine nuclear contribution to $\rho _{n}(T)$ in this case will have the
temperature dependence as in Eq. \ref{fn1} , where $\mu _{n}H$ should be
replaced by the characteristic quadrupole splitting of energy levels.

The influence of the quadrupole nuclear spin splitting on the phase
coherence time $\tau _{\varphi }$ can be the clue to the puzzling difference
between the low temperature dependence of $\tau _{\varphi }$ in $Cu,Au$ and $%
Ag$ wires , observed in \cite{Esteve1}. Indeed, the nuclear spins of \ both $%
Cu$ and $Au$ have a strong quadrupole moment ($s=3/2$) and may act as
inelastic two-level scatterers \cite{ZvDR99},\cite{IFS99}, once their
degeneracy is lifted by the static impurities and other imperfections. The
quadrupole splitting in these materials is known to be of the order of $%
\Delta _{Q}\sim 10^{-3}\div 10^{-2}$ K \cite{WintBk}.

This is not the case for $Ag$ nuclei since their spin is $\ s=\frac{1}{2}$
.In the absence of magnetic Zeeman splitting present, for an electron spin,
just a set of elastic scatterers, and the temperature dependence of $\tau
_{\varphi }$ should obey the s$\tan $dard theory \cite{AAK82}, which indeed
the case in the experiments \cite{Esteve1}.

It is interesting to continue the measurements of $\tau _{\varphi }$ on
other materials with (as $Al:,s=5/2$) \ and without ($Pt$ and $Sn,s=1/2$)\
the quadrupole nuclear spin splitting.

The Kondo effect appears in $\rho _n(T)$ in higher orders of $g$ Fig. 1. In
analogy with the magnetic impurities the first temperature correction to $%
\rho _n(T)$ is 
\begin{equation}
\delta \rho _{1\approx }\rho _{oo}g_n^3\ln \frac{\varepsilon _F}T
\label{deltarho1}
\end{equation}

For positive magnetic nuclear moments one has $g_{n}>0$ and the interaction
between the electron and the nuclear spins favors the antiferromagnetic
ordering of moments $\mu _{e}$ and $\mu _{n}$. This gives the usual Kondo
effect on nuclei, i.e. $\rho (T)$ has a minimum at $T_{o}\approx g^{\frac{3}{%
2}}\varepsilon _{F}$ . For metals with large $Z$: $T_{o}\approx 10^{-3}\div
10^{-2}$ $^{o}K$ . Note, that for the nuclear contribution to $\rho $ there
is no need to summarize the powers of $\ln \frac{\varepsilon _{F}}{T}$ ,
since the nuclear Kondo temperature $T_{k}=\varepsilon _{F}e^{-\frac{1}{g_{n}%
}}$ is so low that the nuclear spin interaction start to play the main role.
These are the direct dipole-dipole interaction between nuclei and their
interaction via conduction electrons. The interaction constant $J^{+}$ is of
the order of $T_{c}^{{}}$ , the temperature of the nuclear ordering. In the
case when the interaction via conduction electrons is stronger than the
direct one , $J^{+}\approx $ $g^{2}\varepsilon _{F}$ . It can be shown that
the contribution to $\rho $ from the nuclei-nuclei interactions 
\begin{equation}
\delta \rho _{\approx }\rho _{oo}g_{n}^{2}\frac{J^{+}}{T}  \label{deltarho3}
\end{equation}
are comparable to these of the electron contribution $\rho _{oo}\left( \frac{%
T}{\varepsilon _{F}}\right) ^{2}$ at $T_{1}\sim \left( J^{+}\varepsilon
_{F}^{2}\right) ^{\frac{1}{3}}g^{\frac{2}{3}}\sim \varepsilon _{F}^{{}}g^{%
\frac{4}{3}}$.

By comparing $T_o$ and $T_1$ one concludes that while considering the Kondo
effect on nuclear spins the nuclear spin interaction should not be
neglected, since the nuclear spin concentration is always of order of unity.
This to be compared with the usual Kondo effect in the case of low
concentration of magnetic impurities $C_m$ , where the interaction between
the localized moments are of the second order with respect to $C_m$. The
correction to $\rho $ , Eq. \ref{fn1} is analogous to the correction $\sim
J^{+}$ to the nuclear susceptibility

\[
\chi _n(T)\sim \frac 1T\left( 1+\frac{J^{+}}T\right) 
\]
By measuring the URR in a metal one can therefore establish the sign of $%
J^{+}$ and to predict the type of the nuclear order at very low temperature.

The high temperature $T\gg J^{+}\approx T_{c }$expansion of the residual
resistivity of a nonmagnetic metal is

\begin{equation}
\frac{\rho (T)}{\rho _{oo}}\approx C_{o}+\left( \frac{T}{\varepsilon _{F}}%
\right) ^{2}+g_{n}^{2}\left( 1+g_{n}\ln \frac{\varepsilon _{F}}{T}+\frac{%
J^{+}}{T}\right)   \label{rho(T)7}
\end{equation}

It follows, from Eq. \ref{rho(T)7} that in metals with large $Z$ the nuclear
effects will contribute to resistivity already at temperatures of the order
of $0.1^{o}K.$ Note that the nuclear contribution is nonanalytical in $T$
and should be compared with the vanishing, at low temperatures, $T^{2}$
contribution rather than with $\rho _{oo}C_{o}$ (see for more details \cite
{DVW96})$.$ 

In conclusion we have suggested that the hyperfine interaction between the
conduction electron spins and nuclear spins may result in universal residual
resistivity in clean metals at low temperatures. Apart of the fundamental
nature of this problem, the natural limitations on the mean free path are
decisive in the semiconductor based high speed electronic devices, like
heterojunctions and quantum wells. We outline, that the nuclear spin
quadrupole splitting due to the static imperfections may be partly
responsible for the low temperature behavior of resistivity in such metals
as $Au$ and $Cu$. This mechanism should not be operative in $Ag$ where the
nuclear spin is $s=1/2$ , in agreement with the recent experimental
observations \cite{DVW96}. We note also that the influence of the nuclear
spins on resistivity should disappear at very low temperatures, where the
nuclear spins magnetically order (see e.g. \cite{DVW96},\cite{RHP97}).

We acknowledge illuminative discussions with B. Altshuler, D. Esteve, T.
Hermannsdorfer,V. Kravtsov, T.Maniv, B. Spivak,  P.C.E.\ Stamp and R. Webb.

\newpage\

\end{document}